# Cooperatively Modulating Magnetic Anisotropy and Colossal Magnetoresistance via Atomic-Scale Buffer Layers in Highly Strained La$_{0.7}$Sr$_{0.3}$MnO$_3$ Films

Sheng Li[1,2], Zengxing Lu[1,2], Bin Lao[1,2], Xuan Zheng[1,2], Guoxin Chen[3], Run-Wei Li[1,2,4*] and Zhiming Wang[1,2,4*]

[1]CAS Key Laboratory of Magnetic Materials and Devices, Ningbo Institute of Materials Technology and Engineering, Chinese Academy of Sciences, Ningbo 315201, China
[2]Zhejiang Province Key Laboratory of Magnetic Materials and Application Technology, Ningbo Institute of Materials Technology and Engineering, Chinese Academy of Sciences, Ningbo 315201, China
[3]Testing Center, Ningbo Institute of Materials Technology and Engineering, Chinese Academy of Sciences, Ningbo 315201, China
[4]Center of Materials Science and Optoelectronics Engineering, University of Chinese Academy of Sciences, Beijing 100049, China
*Email: zhiming.wang@nimte.ac.cn, runweili@nimte.ac.cn

## Abstract

Simultaneous control of magnetic anisotropy and magnetoresistance, especially with atomic scale precision, remains a pivotal challenge for realizing advanced spintronic functionalities. Here we demonstrate cooperative continuous control over both magnetoresistance and magnetic anisotropy in highly strained La$_{0.7}$Sr$_{0.3}$MnO$_3$ (LSMO) thin films. By inserting varying perovskite buffer layers, compressively strained LSMO films transition from a ferromagnetic insulator with out-of-plane magnetic anisotropy to a metallic state with in-plane anisotropy. Atomic-scale buffer layer insertion enables remarkably acute, precise control to sharply modulate this magnetic phase transformation. A gigantic 10,000% modulation of the colossal magnetoresistance (CMR) and an exceptionally sharp transition from out-of-plane to in-plane magnetic anisotropy are attained in just a few contiguous layers. These atomic-scale correlations among electronic, magnetic, and structural order parameters yield flexible multifunctional control promising for next-generation oxide spintronics.

**Introduction**

Spintronic devices represent an emerging technology that harnesses both the charge and spin degrees of freedom in electrons, promising exciting new capabilities beyond conventional electronics[1-3]. Realizing the full potential of spintronics necessitates optimizing the two vital functions—reading and writing data—which impose diverse demands on the magnetic materials[4,5]. The reading function relies on sensitive magnetoresistance effects in response to specific magnetization states. Such as giant magnetoresistance[6,7] and tunneling magnetoresistance[8,9] have been uncovered over the years, leading to commercial applications in magnetic recording and sensing. On the other hand, the writing function requires the ability to manipulate the magnetization direction in data storage elements. This depends critically on the magnetic anisotropy, which determines the stability of the magnetization against thermal fluctuations and applied fields[10]. In addition, the magnetic anisotropy greatly influences overall device performance metrics including scalability, power consumption, speed, and thermal stability[5,11,12]. With continued scaling down of device dimensions, it becomes increasingly difficult to maintain a sufficient anisotropy barrier to prevent spontaneous switching of the magnetization while also keeping the required writing fields accessible[13]. Therefore, simultaneously tuning both magnetic anisotropy along with sensitive magnetoresistance effects is imperative but remains a challenge.

Transition metal oxide $La_{0.7}Sr_{0.3}MnO_3$ (LSMO) materials possess rich and tunable magnetic properties stemming from the interplay of spin, charge and orbital interactions as well as flexible valence states[14-18]. This enables sensitive control over the key properties like magnetoresistance and magnetic anisotropy, making LSMO highly promising for next-generation spintronics[19-21]. For instance, LSMO demonstrates versatile magnetoresistance effects including anisotropic[22], giant[23], tunneling[24] and colossal[25,26] magnetoresistance. Meanwhile, its magnetic anisotropy energy can also be readily engineered using strain, interfacial, and composite approaches[21] Thanks to rapid advances in the atomic-scale precision epitaxial growth of complex oxides, single unit cell layer control has enabled several key demonstrations[27-29]. Via tuning oxygen octahedral rotations at the atomic-scale, the relative strength of magnetic anisotropy in LSMO can be control in the plane[30,31]. Perpendicularly magnetic anisotropy and associated magnetized tunnel junctions have also been engineered by tailoring interfacial effects[32-35]. Furthermore, atomic control of LSMO film thickness has induced a metal-to-insulator transition coupled with large CMR modulation. However, the simultaneous control and enhancement of both magnetic anisotropy and magnetoresistance functionalities at the atomic-scale remain elusive.

In this work, we develop a feasible approach to simultaneously control multiple correlated functional properties in complex oxides at an atomic scale. High quality epitaxial LSMO thin films are grown by pulsed laser deposition (PLD) with controlled layer-by-layer insertion of various perovskite buffer layers ($CaTiO_3$, $SrIrO_3$, *etc.*) on single crystal $LaAlO_3$ (LAO) substrates to impose structural distortions in the LSMO. By judiciously tuning buffer layer thickness and lattice symmetry mismatch at the atomic scale, we demonstrate a cooperative modulation of electronic, magnetic and structural order parameters in the LSMO films. A remarkably sharp metal-insulator transition is concurrently induced along with continuous reorientation of the magnetic easy axis, achieving over two orders of magnitude change in colossal magnetoresistance across nanoscale regions spanning just a few unit cells.

**Results**

High-quality oxide heterostructures composed of 15-unit cells (u.c.) LSMO and 5-u.c. various buffer layers are grown on (001)-oriented LAO substrates by PLD. The buffers used here are $CaTiO_3$ (CTO), $NdGaO_3$ (NGO), $GdScO_3$ (GSO) and $SrIrO_3$ (SIO), all possessing a common orthorhombic crystal structure. Figure 1(a) compares the pseudocubic lattice constants of the components and substrate. The lattice mismatch is calculated by $\varepsilon = \frac{a_{LAO} - a_{film}}{a_{LAO}} \times 100\%$, where $a_{LAO}$ and $a_{film}$ are the lattice constants of substrate and films. With a small 3.79 Å lattice constant, LAO induces -2.24% compressive strain in the bare LSMO films. In contrast, CTO and NGO buffers have smaller mismatch of -0.34% and -1.71%, while GSO and SIO have huge mismatch of -4.35%. To confirm coherent epitaxial growth despite the huge mismatch, the lattice structure of SIO/LSMO bilayer is systematically characterized by XRD (Fig. 1(b)) and STEM (Fig. 1(c)) measurements. The XRD $\theta$-$2\theta$ scan reveals the out-of-plane lattice constant of 3.99 Å for both LSMO ($c_{LSMO}$) and SIO/LSMO ($c_{SIO/LSMO}$) films. Besides, the reciprocal space mapping shows that the insertion of 5 u.c. SIO can still preserve the fully strained nature of LSMO (see Supplementary Information Fig. S1 (a)). Moreover, the STEM results directly visualize the epitaxial relation between the film and substrate with sharp interfaces between different layers. These structural analyses demonstrate that the SIO/LSMO heterostructure preserves the fully strained nature.

To examine the impact of the buffer layers on modulating the electronic properties of LSMO films, transport measurements are performed on LSMO heterostructures with varied buffer layers. Figure 1(d) displays the resultant temperature-dependent resistivity curves ($\rho_{xx}(T)$). The trends in resistivity over temperature reveal the metallic or insulating nature of these heterostructures. The bare LSMO film shows rapidly increasing resistivity with decreasing temperature, indicating an insulating state contrary to the metallic bulk LSMO. This metal-insulator transition (MIT) is consistent to the previous reports attributing it to orbital reconstruction induced by high compressive strain from the LAO substrate[36-38]. Heterostructures with CTO and NGO buffer layers exhibit similar insulating behavior. However, GSO- and SIO-buffered LSMO heterostructures display an initial increase and subsequent decrease in resistivity below 298K and 308K respectively, signaling emergence of metallicity. These results indicate that there is a buffer layer variety driven MIT occurred in the LSMO films grown on LAO. Despite the insulating nature of GSO similar to CTO and NGO, the insulating nature of SIO is also confirmed by the temperature-dependent $\rho_{xx}$ of the 8-u.c. SIO/LAO heterostructure shown in Supplementary Information Fig. S2 (a). Overall, the resistivity behavior variation confirms modulation of LSMO's electrical transport properties via judiciously altering buffer layers while preserving coherent epitaxial strain.

To achieve even finer control over the electrical and magnetic properties of strained LSMO, we utilized the tunability of SIO by systematically varying its thickness as an additional degree of freedom. By modulating the SIO buffer layer thickness from 0 to 8 unit cells before growing LSMO films, we could manipulate the insulator-to-metal transition with atomic precision and gain deeper insights into its evolution. Figure 2(a) shows the temperature-dependent electrical resistivity of SIO($n$)/LSMO(15) heterostructures, with the SIO buffer layer thickness ($n$) systematically varied from 0 to 8 u.c., measured over the range of 10-360 K. The overall trend reveals a decrease in resistivity across the entire temperature range as the SIO thickness increases. Remarkably, by incrementally inserting the SIO buffer layer unit cell by unit cell, we induce a transformation of the LSMO films from insulating to metallic behavior between low temperature and the Curie temperature $T_C$, as determined from magnetic characterization shown in Fig. 4(e). Examining Fig. 2(a) in more detail, we observe fully insulating behavior only for the bare LSMO(15) film without a buffer layer. In contrast, even with just a single u.c. of SIO, a temperature-driven MIT emerges in the LSMO. The metal-insulator transition temperature ($T_{MIT}$), marked by arrows, increases from 82 K for 1 u.c. to 320 K for 8 u.c. of SIO as the buffer thickness is incrementally increased.

To illustrate the origin and evolution of the emergent insulating state above the $T_{MIT}$, we have analyzed the temperature-dependent resistivity data using Mott variable-range-hopping (Mott-VRH) models for thermal activated conduction in disordered, localized systems[39-42]. Figure 2(b) depicts the linear fitting with the natural logarithm of resistivity $\ln(\rho_{xx})$ versus $T^{-1/4}$ in the insulating regime for the SIO($n$)/LSMO(15) heterostructures. The Mott-VRH model relates the resistivity $\rho_{xx}(T)$ to a characteristic temperature $T_0$ and a dimension-dependent coefficient $d$ through the expression:

$$\rho_{xx}(T) = \rho_0 \exp(T_0/T)^{[1/(d+1)]}$$

where $\rho_0$ is the resistivity coefficient, and $d$ takes a value of 3 for the three-dimensional systems studied. From the fitted $T_0$ values, we can calculate the Mott hopping activation

energy ($E_M$) using:

$$E_M = 1/4 k_B T (T_0/T)^{1/4}$$

where $k_B$ is the Boltzmann constant. As shown in Fig. 2(c), the extracted $E_M$ decreases from 108.4 meV for bare LSMO to 11.8 meV for 3 u.c. of SIO buffered heterostructure at 300 K. The decreasing $E_M$ with increasing buffer layer thickness indicates that the initial strong carrier localization in the bare LSMO film gradually weakens as more SIO layers are introduced. At critical buffer thicknesses $n = 3$, the transport properties deviate from the strong localization behavior described by the Mott-VRH model. This analysis reveals that carrier localization serves as a key driving force for the metal-insulator transition observed in these heterostructures. The insulating ground state and carrier localization observed in our highly strained LSMO films bears similarities to the behavior reported in other manganite systems such as $La_{1-x}Ca_xMnO_3$ (LCMO) and $Pr_{1-x}Ca_xMnO_3$ (PCMO) thin films[43-48]. These doped manganites also exhibit insulating behavior at low temperatures due to strong electron localization, particularly at lower doping levels. The metal-insulator transition in LCMO and PCMO has been attributed to the interplay between electron-lattice and electron-electron interactions leading to polaronic charge carriers. Our analysis using the Mott-VRH model suggests an analogous role of localization effects in driving the insulating state in strained LSMO films as well. However, in contrast to the dopant-controlled localization in LCMO/PCMO, our work demonstrates the ability to continuously tune the localization and resulting insulating behavior in LSMO using precise structural modulations at the atomic scale through the SIO buffer layer thickness.

To further understand the atomic-scale modulation of transport properties in the SIO($n$)/LSMO(15) heterostructures, we have measured the temperature dependence of resistance under different magnetic fields $H$. Figures 3(a) and 3(b) show the temperature-dependent electrical resistance of LSMO (15) and SIO(8)/LSMO(15) under different $H$ of 0, 3, 6 and 9 T measured in the temperature range from 10 to 360 K. The LSMO(15) film exhibits predominantly insulating behavior across the entire temperature range below 3 T. Remarkably, a metallic phase is induced at higher applied fields of 6 and 9 T, and the $T_{MIT}$ increases with increasing magnetic field strength. This field-induced insulator-to-metal transition is consistent with previous reports on other insulating manganite systems, where applied magnetic fields can disrupt the insulating ground state and drive an insulator-to-metal transition[49]. In contrast, the SIO(8)/LSMO(15) heterostructure displays metallic transport behavior under all magnetic field. However, compared to the only 12% change in resistance between 0 and 9 T in the test temperature range, the $T_{MIT}$ increases with fields up to 3 T but extends beyond our measurement range as the field is further raised from 3 to 9 T. This behavior is similar to that observed in LCMO, arising from the destabilization of the insulating state due to the gradual breakdown of magnetic order above the $T_C$[43]. To quantify the evolution of the magnetoresistive response, we measured the temperature-dependent magnetoresistance curves at $H$ = 0, 3, 6, and 9 T for the full series of heterostructures. The CMR is calculated using the formula [25]:

$$CMR = \frac{\rho_{xx}(0\ T) - \rho_{xx}(H)}{\rho_{xx}(H)}$$

Figure 3(c) shows the CMR of the SIO($n$)/LSMO(15) with different buffer layer thickness at $H$ = 9 T. Overall, the maximum CMR value decreases with increasing SIO thickness. For the SIO buffer layer thickness is 0 and 1 u.c., the CMR initially increases with decreasing temperature before reaching a peak, then rapidly drops. The maximum CMR of 12179 % is observed in LSMO (15) at 94 K. The corresponding temperature of the CMR maximum peak, denoted as $T_{CMR}$, does not follow a monotonic trend initially. However, as the SIO thickness increases beyond 2 u.c., $T_{CMR}$ increases with buffer layer thickness as shown in the Fig. 3(d).

To directly demonstrate the correlation between magnetic ordering and underlying electronic and structural changes, we characterized the magnetic properties of the SIO/LSMO heterostructures and tracked their evolution as a function of buffer layer thickness. Figures 4(a) and 4(b) show the magnetic hysteresis loops $M(H)$ for the LSMO (15) and SIO(8)/LSMO (15) films measured in-plane and out-of-plane at 10 K, respectively. Due to the orbital reconstruction and charge ordering stabilized by the large compressive strain, the bare LSMO (15) film exhibits detectable perpendicular magnetic anisotropy (PMA), consistent with previous reports[38]. In contrast, the SIO(8)/LSMO (15) heterostructure displays in-plane magnetic anisotropy (IMA). To systematically investigate this anisotropy transition, we measured the in-plane and out-of-plane $M(H)$ for the complete series of heterostructure with varying SIO thickness. From these measurements, we calculated the effective magnetic anisotropic energy (MAE) $K_{eff}$. As shown in Fig. 4 (e), a sharp transition in the magnetic easy axis orientation is observed with increasing buffer layer thickness around 2-3 u.c.. The easy axis reorients from out-of-plane to in-plane, characterized by a decrement in MAE from +2.69 to -1.06 $\times 10^6$ erg/cm$^3$. The ability to continuously modulate the MAE in LSMO is demonstrated through this atomic layer-by-layer insertion of the SIO buffer.

The temperature-dependent magnetization $M(T)$ for LSMO(15) and SIO(8)/LSMO(15) measured under magnetic field of 0.05 T applied along in-plane and out-of-plane is shown in Figs. 4(c) and 4(d). The magnetization decreases smoothly upon heating and vanishes after reaching the $T_C$, which is determined from the intersection of the tangent line near the phase transition point with the temperature axis (plotted inset). The $T_C$ for LSMO (15) is only 203 K but gradually increased to 308 K as the SIO buffer is inserted layer by layer up to 8 u.c., as summarized in Fig. 4(e). Moreover, the magnetic moment for the bare LSMO film remains dominant along with the out-of-plane direction, exhibiting a larger value than that along the in-plane direction. Near the critical SIO thickness $n = 3$, competition between out-of-plane and in-plane moments is observed across the transition temperature range (see Supplementary Fig. S3(a) and S3(b)). For the other heterostructures, the direction of easy axis persists domination over the entire temperature range. Finally, we estimated the saturation magnetization $M_S$ for different heterostructures at 10 K as shown in Fig. 4(f), which shows a minimum of 2.48 $\mu_B$/Mn around 2 u. c. SIO thickness and a maximum of 3.89 $\mu_B$/Mn for 8 u.c. SIO. These results on the evolution of magnetic anisotropy, $T_C$, and $M_S$ with increasing SIO insertion point to an intimate coupling between the magnetic properties and structural/electronic degrees of freedom in the highly strained SIO/LSMO heterostructures.

**Discussion**

The layer-by-layer insertion of perovskite buffer layers in LSMO thin films enables an unprecedented ability to continuously modulate the CMR and magnetic anisotropy simultaneously at an atomic-scale precision. The insertion of just 2-3 u.c. of SIO drives a sharp transition in the magnetic easy axis from a perpendicular orientation in bare LSMO films to in-plane direction. Concurrently, the CMR exhibited a systematic evolution - transitioning from large values in the insulating bare LSMO to a weaker response as metallicity was induced with increasing SIO thickness. This capability to manipulate the CMR and magnetic anisotropy in a concerted manner by atomic layer control highlights the intricate coupling between spin, charge, orbital and lattice degrees of freedom in these complex oxide systems.

In a bare LSMO films grown on LAO, the large compressive epitaxial strain stabilizes a ferromagnetic insulating state that deviates from the metallic ferromagnetism displayed in bulk LSMO. This metallicity in unstrained LSMO arises from double exchange interactions whereby alignment of Mn $t_{2g}$ spins enables hopping of $e_g$ electrons[16-18]. Similar strain-enabled insulation has been reported even in otherwise conducting LCMO membranes[43-45]. Intriguingly, the emergence of these insulating states closely resembles the colossal magnetoresistance (CMR) properties of doped manganites like PCMO, where magnetic field-tuning is known to drive an insulator-metal transition[46,50]. The insulating ground state likely originates from strain-enhanced cooperative Jahn-Teller distortions which could promote orbital-ordering of $e_g$ electron states, leading to formation of localized spin-charge-orbital ordered regions embedded within the conducting ferromagnetic matrix[25].

However, unlike previous works that rely on conventional approach like change composition, chemical doping, applied fields, and carrier concentrations to explore CMR by disrupting the delicate balance of spin, charge and orbital interactions[25,26,51]. Our work demonstrates a new route to continuously tune the CMR by controlling the degree of Jahn-Teller distortion systematically. Specifically, the insertion of the SIO buffer layers can alleviate epitaxial strain effect in LSMO films via octahedral rotations and distortions. This suppresses the cooperative orbital ordering caused by Jahn-Teller distortions, thereby inducing a MIT seen from transport measurements. We achieve continuous tuning of magnetoresistance over a range of 10000% with concomitant switching from an insulating state to a metallic regime upon increasing SIO layer thickness. Moreover, the intricate correlations between the various degrees of freedom are further exemplified by the concomitant transition in the magnetic easy axis from out-of-plane to in-plane direction. The ability to manipulate these disparate properties in a concerted manner, enabled by atomic layer-by-layer synthesis, underscores the delicate interplay between spin, charge, orbital and lattice governing the rich phenomenology of these complex oxide heterostructures. It paves exciting pathways to leverage such strong couplings between multiple quantum mechanical variables for next-generation electronics.

**Conclusion**

To conclude, we have demonstrated an unprecedented capability of simultaneously modulating the colossal magnetoresistance and magnetic anisotropy in highly strained LSMO thin films at an atomic-scale precision through systematic insertion of the oxide buffer layers. The layer-by-layer SIO insertion enables continuous tuning of the metal-insulator transition, gradually reducing the carrier localization and inducing metallicity in originally insulating LSMO films under compressive strain. Concurrently, the magnetic easy axis rotates sharply from an out-of-plane orientation in bare LSMO films towards an in-plane anisotropy above 2-3 unit cells of inserted SIO. The CMR evolves in tandem, transitioning from very high values in the insulating regime to a weaker response in the metallic state. These effects originate from interfacial octahedral rotations which control the orbital reconstruction and Jahn-Teller distortions in LSMO. Our work puts forth oxide heterostructures as a versatile platform to unravel intricate couplings between spin, charge, orbital and lattice degrees of freedom in complex oxides. The capability to manipulate electronic and magnetic properties simultaneously opens avenues to explore emergent intertwined quantum states. Harnessing this tunable entanglement of properties by atomic-scale control charts a roadmap to develop novel multifunctional electronics and spintronics leveraging correlated oxide materials.

## Method

### Film Fabrication and Structure characterization

The LSMO/buffer layer heterostructures are fabricated using pulsed laser deposition (PLD). The films are deposited on LAO substrates layer-by-layer using a KrF excimer laser ($\lambda$ = 248 nm) under the following conditions: temperature of 700°C, laser energy density of 2.0 J/cm$^2$, pulse frequency 2 Hz, oxygen partial pressure 0.12 mbar. In-situ monitoring by high pressure reflective high energy electron diffraction (RHEED) enables precision tracking of growth mode and film thickness. The crystalline quality and epitaxial strain state of the films are examined by high-resolution X-ray diffraction (HRXRD). Interfacial atomic structure is investigated using scanning transmission electron microscopy (STEM).

### Device fabrication, Magnetic & Electrical measurements

Electrical transport properties are measured on square 4.4 × 4.4 mm$^2$ LSMO heterostructures with Ti/Au electrodes at the corners deposited by electron-beam evaporation, using van der Pauw method. Temperature and magnetic field dependent resistivity and magnetoresistance are measured using a home-built system. Magnetization data is collected with a SQUID magnetometer from Quantum Design.

## Acknowledgments

This work was supported by the National Key Research and Development Program of China (No. 2019YFA0307800), the National Natural Science Foundation of China (Nos. 12174406, 11874367, 51931011, 52127803), K.C.Wong Education Foundation (GJTD-2020-11), the Ningbo Key Scientific and Technological Project (Grant No. 2022Z094), the Ningbo Natural Science Foundation (No. 2023J411), the Natural Science Foundation of Zhejiang province of China (No. LR20A040001) and "Pioneer" and

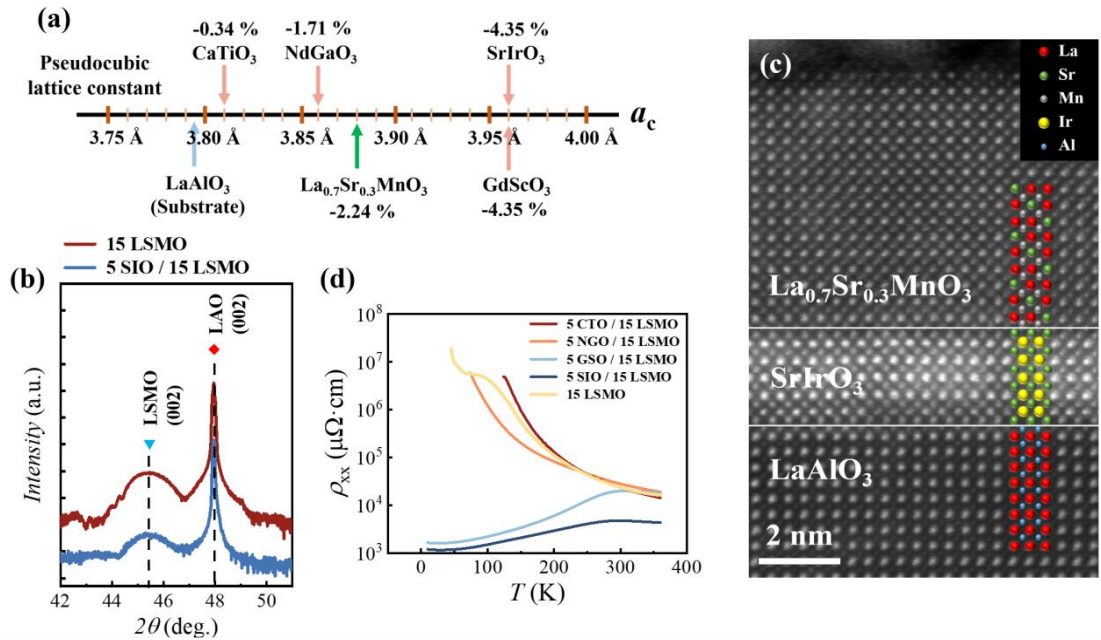

**FIG. 1. Structural characterization and transport properties of highly strained LSMO films with varying buffer layers.** (a) Comparison of pseudocubic lattice constant of LSMO films (green arrow), CTO, NGO, GSO and SIO buffer layers (orange arrows), and LAO substrate (black arrow). Percentages indicate the lattice mismatch between each layer and LAO. (b) XRD $\theta$-$2\theta$ scan around (002) peaks for a LSMO film and a SIO/LSMO bilayer grown on LAO substrates. (c) High-resolution STEM image captured along the [100] zone axis, showing the interface between the SIO buffer layer and the LSMO film. (d) Temperature-dependent resistivity $\rho_{xx}$ ($T$) for LSMO films grown with various 5-u.c. buffer layers and without a buffer layer.

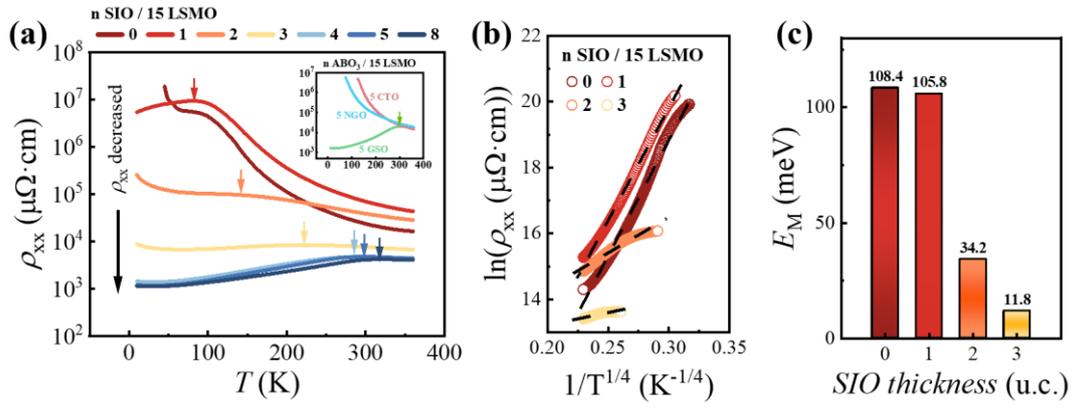

**FIG. 2. Transport properties and Mott variable-range hopping analysis of SIO(*n*)/LSMO(15) heterostructures.** (a) Temperature-dependent resistivity $\rho_{xx}$ for heterostructures with varying SIO buffer layer thickness $n = 0 – 8$ u.c., measured over 10-360 K. A systematic decrease in resistivity across the temperature range is observed with increasing $n$. Arrows mark the metal-insulator transition temperature $T_{MIT}$ which increases from 82 K to 320 K as n increases from 1 to 8 u.c.. Inset shows $\rho_{xx}(T)$ curves for films with other buffer layers. (b) Logarithm of $\rho_{xx}$ plotted versus $1/T^{1/4}$ in the insulating regime for different SIO(*n*)/LSMO(15) samples, fitted using the three-dimensional Mott variable-range hopping model (dashed black lines). (c) Calculated Mott hopping energy $E_M$ decreases with increasing SIO buffer thickness, extracted from the fitted characteristic temperatures $T_0$ in (b).

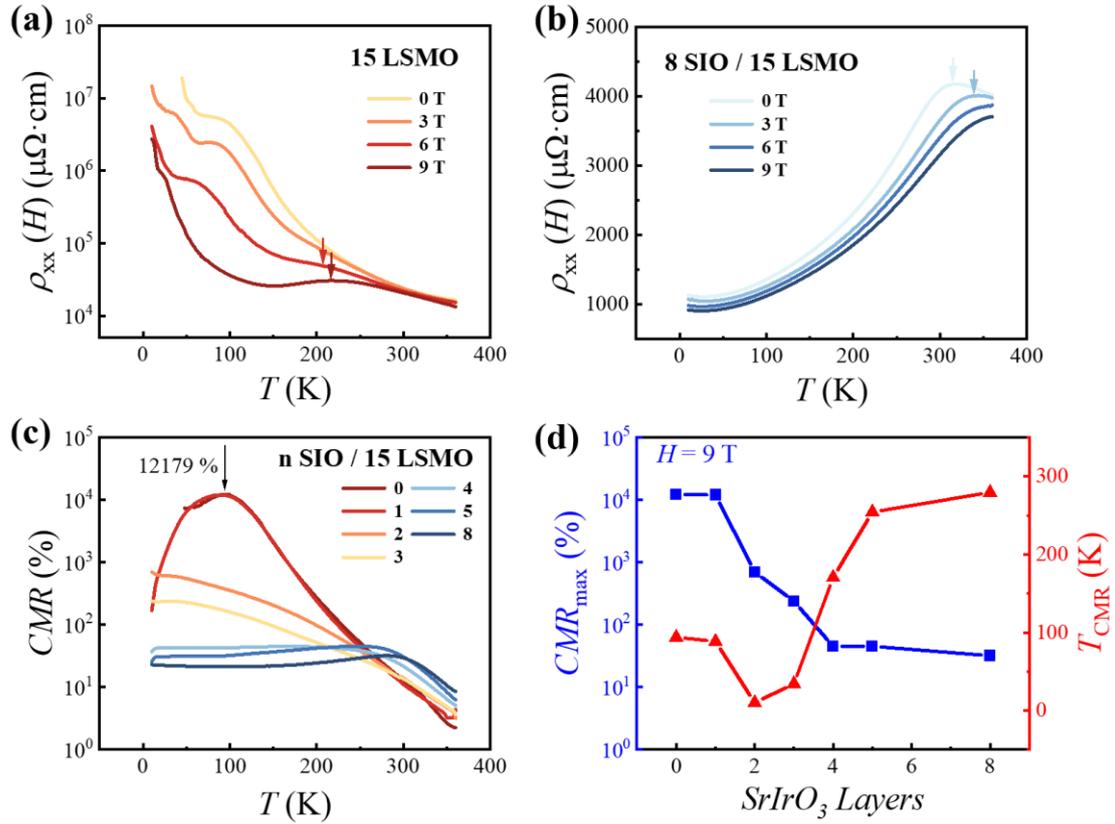

**FIG. 3. Colossal magnetoresistance in SIO (*n*)/LSMO (15) heterostructures.** (a-b) temperature-dependence of the electrical resistance $\rho_{xx}$ under different applied magnetic fields for LSMO(15) (a) and SIO(8)/LSMO(15) (b) measured from 10–360 K. (c) Colossal magnetoresistance for heterostructures with varying SIO buffer thickness *n* at $H = 9$ T. (d) Maximum CMR value and the corresponding temperature $T_{CMR}$ plotted as a function of SIO buffer layer thickness.

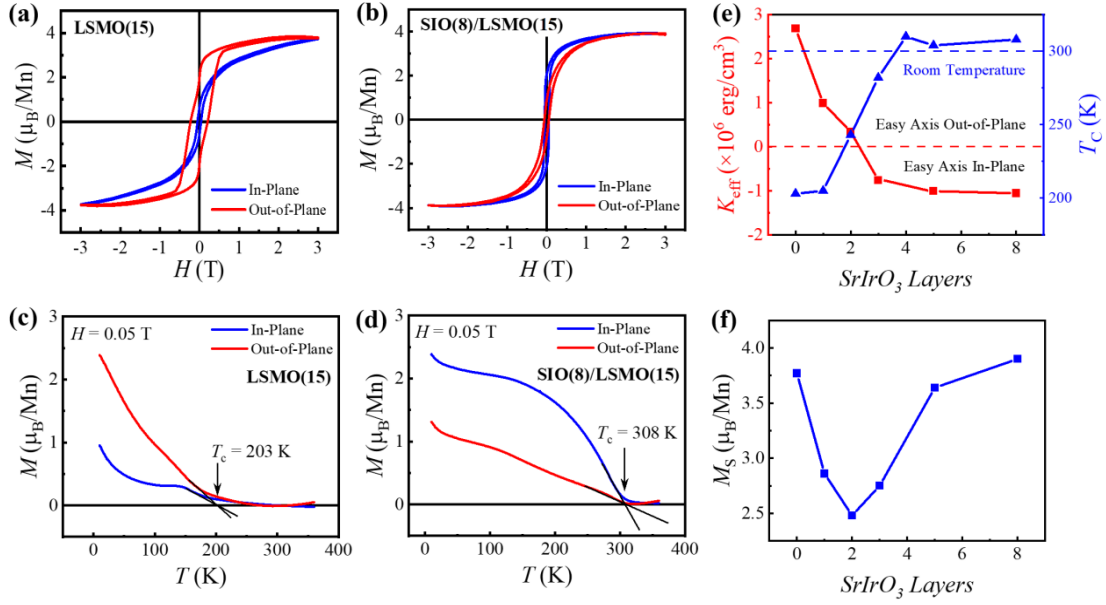

**FIG. 4. Magnetic properties of SIO(*n*)/LSMO(15) heterostructures.** (a-b) In-plane and out-of-plane magnetization hysteresis loops of LSMO(15) and SIO(8)/LSMO(15) measured at 10 K, respectively. (c-d) Corresponding in-plane and out-of-plane magnetization vs temperature curves measured at $H$ = 0.05 T. $T_C$ is determined by the intersection points of $M(T)$ curve's tangent and the $M = 0$ axis. (e) Effective magnetic anisotropic energy (MAE) $K_{eff}$ and $T_C$ as a function of the SIO buffer layer thickness. (f) Saturation magnetic moment $M_s$ at 10 K versus SIO thickness.

# Supplementary Materials

# Cooperatively Modulating Magnetic Anisotropy and Colossal Magnetoresistance via Atomic-Scale Buffer Layers in Highly Strained $La_{0.7}Sr_{0.3}MnO_3$ Films

Sheng Li[1,2], Zengxing Lu[1,2], Bin Lao[1,2], Xuan Zheng[1,2], Guoxin Chen[3], Run-Wei Li[1,2,4*] and Zhiming Wang[1,2,4*]

[1]CAS Key Laboratory of Magnetic Materials and Devices, Ningbo Institute of Materials Technology and Engineering, Chinese Academy of Sciences, Ningbo 315201, China
[2]Zhejiang Province Key Laboratory of Magnetic Materials and Application Technology, Ningbo Institute of Materials Technology and Engineering, Chinese Academy of Sciences, Ningbo 315201, China
[3]Testing Center, Ningbo Institute of Materials Technology and Engineering, Chinese Academy of Sciences, Ningbo 315201, China
[4]Center of Materials Science and Optoelectronics Engineering, University of Chinese Academy of Sciences, Beijing 100049, China
*Email: zhiming.wang@nimte.ac.cn  runweili@nimte.ac.cn

## S1. Structure measurement for SrIrO$_3$($n$)/La$_{0.7}$Sr$_{0.3}$MnO$_3$ (15)

In order to ensure that the SIO($n$)/LSMO(15) heterostructure can still maintain the epitaxial characteristics and original structural phase on the LaAlO$_3$(001) substrate with huge compressive strain, we performed high-resolution XRD scanning of the SIO/LSMO bilayers.

Figure S1(a) exhibits the epitaxial relationship between the SIO($n$)/LSMO(15) film and LAO(001) substrate characterized via X-ray reciprocal space mapping (RSM) along (-103) direction, in which the identical $h$ values strongly suggest that the film is fully strained by the substrate. Figure S1(b) shows the $\theta$-$2\theta$ scan of a series of SIO(n)/LSMO(15) film, in which distinct (002) peaks of both film and substrate implying the epitaxial growth of the film. According to the peak position, the out-of-plane lattice constant $c_{LSMO}$ is calculated as 3.99 Å by Bragg's Law, which is larger than the bulk value 3.88 Å. The $c$/$a$ ratio is 1.052 indicating an in-plane compressive strain in the deposited film.

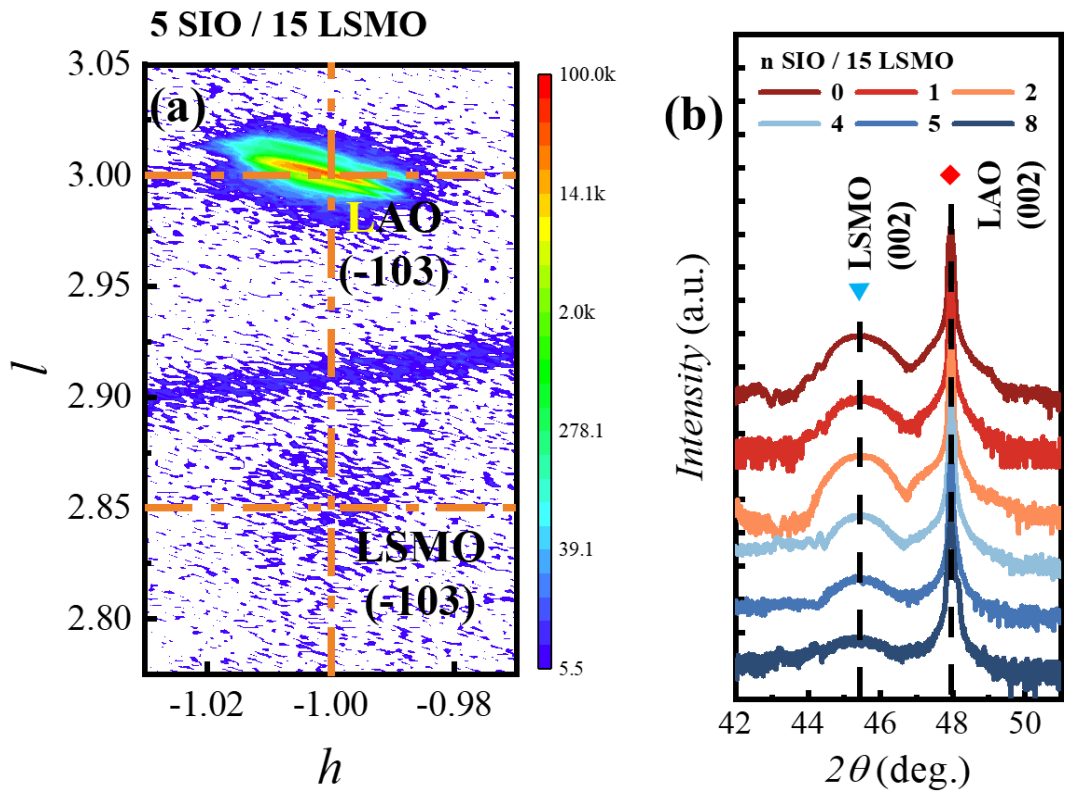

**FIG. S1 a** The results of reciprocal space mapping around (-103) peak for the prepared

SIO(5)/LSMO(15) film on LAO(001) substrate. **b** XRD θ-2θ scans for the series of SIO($n$)/LSMO(15) ($n$ = 0-5, 8) heterostructure on the LAO (001) substrate. Inverted triangle and rhombus represent the (002) peak of LSMO and LAO respectively.

## S2. Temperature-dependent resistivity curve of SIO on LAO(001) substrate

In order to confirm that the metallicity in the SIO/LSMO bilayer is caused by the whole film, rather than the semi-metallic properties of the SIO or the LSMO itself. SIO(8) and SIO(8)/LSMO(15) samples are deposited on the LAO substrate as controls and the resistivity temperature-dependent curves of the LSMO(15) are compared. Both LSMO (15) and SIO(8) exhibit significant insulator behavior which is consistent with the references [1].

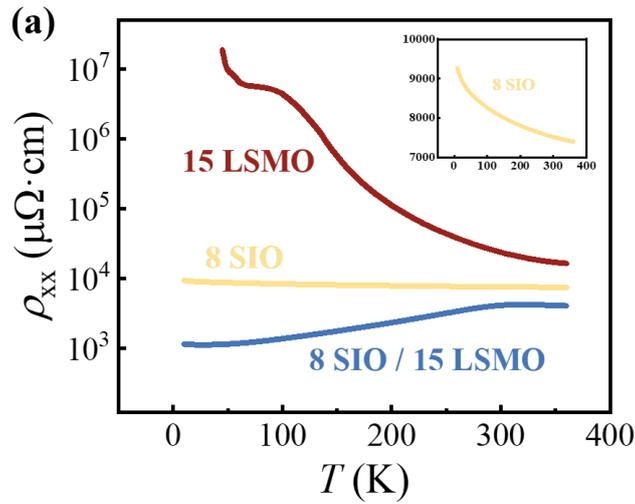

**FIG S2. a** Resistivity vs temperature of LSMO(15), SIO(8) and SIO(8)/LSMO(15) films. Inset shows the zoom-in temperature-dependent resistivity of SIO(8) films grown on the LAO(001) substrate.

## S3. Competition between in-plane and out-of-plane moment with critical SIO thickness

With the increasing thickness of SIO, the easy axis of SIO/LSMO also changes from out-of-plane to in-plane. At the extreme SIO thicknesses, the direction of the easy axis, which is dominant in the hysteresis loop, can also dominate in the curve of M-T during the entire temperature range according to the Fig S3(a)&(c). However, at the critical SIO thickness, $n = 3$, the dominant relationship between the out-of-plane and in-plane is not stable and they only dominate in specific temperature ranges as shown in the Fig S3(b). In the Fig S3(d), the hysteresis loop also indicates that the easy axis distinction between the in-plane and out-of-plane is blurred.

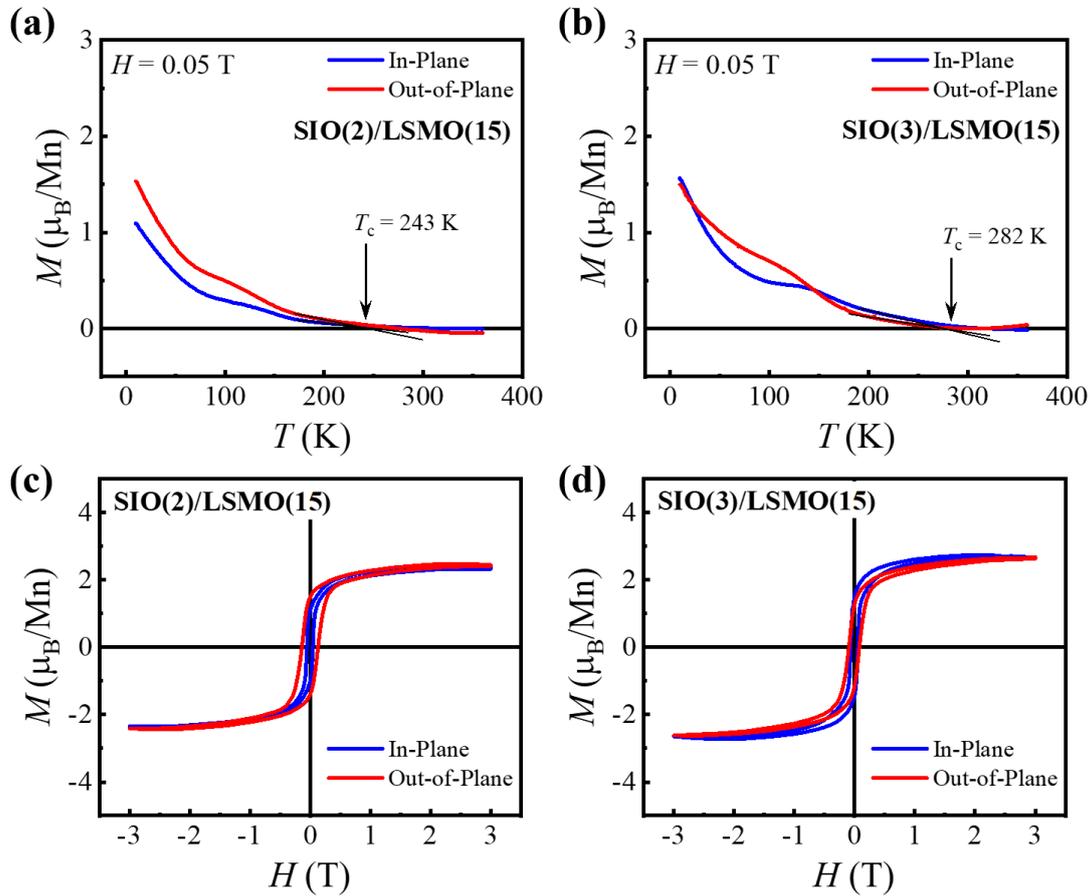

**FIG S3. a-b** In-plane and out-of-plane magnetization vs temperature curves of SIO(2)/LSMO(15) and SIO(3)/LSMO(15) measured at $H = 0.05$ T. **c-d** Corresponding in-plane and out-of-plane magnetization hysteresis loops measured at 10 K.